\documentclass[conference]{IEEEtran}
\makeatletter
\def\ps@IEEEtitlepagestyle{%
  \def\@oddfoot{\mycopyrightnotice}%
  \def\@oddhead{\hbox{}\@IEEEheaderstyle\leftmark\hfil\thepage}\relax
  \def\@evenhead{\@IEEEheaderstyle\thepage\hfil\leftmark\hbox{}}\relax
  \def\@evenfoot{}%
}
\def\mycopyrightnotice{%
  \begin{minipage}{\textwidth}
  \scriptsize
  \copyright~2021 IEEE. Personal use of this material is permitted. Permission from IEEE must be obtained for all other uses, in any current or future media, including reprinting/republishing this material for advertising or promotional purposes, creating new collective works, for resale or redistribution to servers or lists, or reuse of any copyrighted component of this work in other works. 
  
  This work has been accepted at the Design Automation Conference (DAC’21).
  \end{minipage}
}
\makeatother

\usepackage[backend=biber,style=ieee,natbib=true]{biblatex} 
\usepackage{amsmath,amssymb,amsfonts}
\usepackage{graphicx}
\usepackage{textcomp}
\usepackage{xcolor}

\usepackage{booktabs}
\usepackage[binary-units=true]{siunitx}
\usepackage{array,multirow}
\usepackage{hyperref}
\usepackage{amsmath}
\usepackage{enumitem}
\usepackage{algorithm}
\usepackage{subfigure}
\usepackage{graphicx}
\usepackage{tablefootnote}
\usepackage[para]{footmisc}

\usepackage[noend]{algpseudocode}

\newcommand{\figref}[1]{Figure~\ref{#1}}
\newcommand{\secref}[1]{Section~\ref{#1}}

\newcommand{\tabref}[1]{Table~\ref{#1}}

\newcolumntype{L}[1]{>{\raggedright\let\newline\\\arraybackslash\hspace{0pt}}m{#1}}
\newcolumntype{C}[1]{>{\centering\let\newline\\\arraybackslash\hspace{0pt}}m{#1}}
\newcolumntype{R}[1]{>{\raggedleft\let\newline\\\arraybackslash\hspace{0pt}}m{#1}}

\ifCLASSINFOpdf
\else
\fi
\begin{document}
%
\title{High-Performance FPGA-based Accelerator for Bayesian Neural Networks}


%
\author{\IEEEauthorblockN{Hongxiang Fan\IEEEauthorrefmark{1}\IEEEauthorrefmark{3}\IEEEauthorrefmark{5},
Martin Ferianc\IEEEauthorrefmark{2}\IEEEauthorrefmark{3},
Miguel Rodrigues\IEEEauthorrefmark{2}, 
Hongyu Zhou\IEEEauthorrefmark{6},
Xinyu Niu\IEEEauthorrefmark{4} and
Wayne Luk\IEEEauthorrefmark{1}}
\IEEEauthorblockA{\IEEEauthorrefmark{1}Department of Computing, Imperial College London, London UK, \textit{\{h.fan17, w.luk\}@imperial.ac.uk}}
\IEEEauthorblockA{\IEEEauthorrefmark{2}Department of Electronic and Electrical Engineering, University College London, London UK,\\ \textit{\{martin.ferianc.19, m.rodrigues\}@ucl.ac.uk}}
\IEEEauthorblockA{\IEEEauthorrefmark{6}\textit{hongyu.hyzhou@gmail.com}}
\IEEEauthorblockA{\IEEEauthorrefmark{4}Corerain Technologies Ltd., Shenzhen China, \textit{xinyu.niu@corerain.com}}}


\maketitle
\begingroup\renewcommand\thefootnote{\IEEEauthorrefmark{3}}
\footnotetext{Equal contribution.}
\endgroup
\begingroup\renewcommand\thefootnote{\IEEEauthorrefmark{5}}
\footnotetext{Corresponding author.}
\begin{abstract}
Neural networks (NNs) have demonstrated their potential in a wide range of applications such as image recognition, decision making or recommendation systems. 
However, standard NNs are unable to capture their model uncertainty which is crucial for many safety-critical applications including healthcare and autonomous vehicles.
In comparison, Bayesian neural networks (BNNs) are able to express uncertainty in their prediction via a mathematical grounding.
Nevertheless, BNNs have not been as widely used in industrial practice, mainly because of their expensive computational cost and limited hardware performance. 
This work proposes a novel FPGA-based hardware architecture to accelerate BNNs inferred through Monte Carlo Dropout.
Compared with other state-of-the-art BNN accelerators, the proposed accelerator can achieve up to 4 times higher energy efficiency and 9 times better compute efficiency.
Considering partial Bayesian inference,
an automatic framework is proposed,
which explores the trade-off between hardware and algorithmic performance.
Extensive experiments are conducted to demonstrate that our proposed framework can effectively find the optimal points in the design space.
\end{abstract}

%
\IEEEpeerreviewmaketitle

\section{Introduction}\label{sec:introduction}
In recent years, neural networks (NNs) have demonstrated their outstanding performance in a variety of applications ranging from image classification~\citep{ferianc2020vinnas} or segmentation~\citep{kendall2015bayesian} to human action recognition~\citep{fan2019f}. However, one of the main drawbacks in standard NNs is that they are not able to capture the model uncertainty which is crucial for many safety-critical applications such as healthcare~\citep{liang2018bayesian} or autonomous vehicles~\citep{azevedo2020stochasticyolo}.
In contrast to standard NNs, Bayesian neural networks (BNNs)~\cite{neal1993bayesian}, which adopt Bayesian inference to provide a principled uncertainty estimation, have become more popular in these applications.

BNNs~\citep{neal1993bayesian} can describe complex stochastic patterns with well-calibrated confidence estimates. An example of this is shown in Figure~\ref{fig:comparison}, which demonstrates that the BNN is uncertain in its predictions when shown completely irrelevant input, in comparison to a standard NN, which is wrongfully overconfident. Hence with BNNs, we can treat special cases explicitly~\citep{gal2016dropout} and they have become relevant in applications where the notion of uncertainty is essential.

However, the advantage of BNNs comes with a burden: due to the high dimensionality of modern BNNs, it is intractable to analytically compute their predictive uncertainty.
Instead, it is necessary to approximate the predictive distribution through Monte Carlo sampling that requires the users to perform repeated sampling of random numbers and then run the same input data through the BNN multiple times, which degradates the hardware performance.
Several algorithmic approximation techniques and hardware architectures~\citep{cai2018vibnn, myojin, 9116302} have been proposed to improve the hardware performance of BNNs. Nevertheless, there are particular drawbacks in these approaches: \textit{1)} The implementation needs of both an NN engine and a sampler makes the design resource and memory-demanding,
and thus current accelerators can only support BNNs consisting solely of linear layers or binary operations, which does not reflect the need in the industrial or research communities in terms of the current state-of-the-art BNN architectures; \textit{2)} To obtain the uncertainty prediction, these accelerators simply perform $S$ forward passes through the whole network repeatedly without considering the actual algorithmic needs of BNNs, which makes them $S$ times slower than standard NNs.



\begin{figure}[t]
\centering
\includegraphics[width=0.7\linewidth]{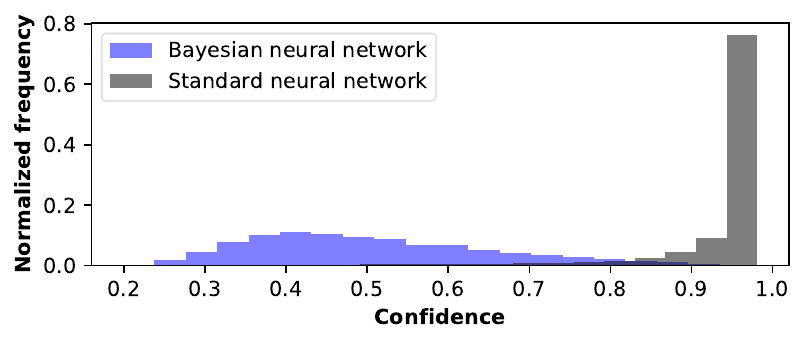}
\caption{Comparison of output confidence histograms for random noise input.}
\label{fig:comparison}
\end{figure}



To address these challenges, we propose an FPGA-based design with the support for fine-grained parallelism to accelerate BNNs inferred through Monte Carlo Dropout (MCD)~\citep{gal2016dropout} with high performance.
The proposed accelerator is versatile to support a variety of BNN architectures.
To further improve the hardware performance, 
we consider partial BNNs to decrease the amount of computation required by BNNs.
An automatic framework is proposed to explore the trade-off between hardware and algorithmic performance, which is able to find a suitable hardware configuration and algorithmic parameters given users' hardware constraints and algorithmic requirements.
In summary, our contributions include:

\begin{itemize}[leftmargin=*]
  \item A novel hardware architecture with an intermediate-layer caching technique to accelerate Bayesian neural networks inferred through Monte Carlo Dropout, which achieves high performance and resource efficiency~(\secref{sec:hardware}).
  \item An exploration framework for hardware-algorithmic performance trade-off and uncertainty estimation provided by partial Bayesian neural network design~(\secref{sec:framework}). 
  \item A comprehensive evaluation of algorithmic and hardware performance on different datasets with respect to different state-of-the-art neural architectures~(\secref{sec:experiments}).
\end{itemize}
\color{black}

\section{Related Work}\label{sec:related_work}

\subsection{Field Programmable Gate Array-based Accelerators}
Acceleration of standard NNs has enjoyed extensive research and industrial interests in the recent years~\citep{mittal2020survey}. Given the high computational demands of NNs, custom hardware accelerators are vital for boosting their performance. The high energy efficiency, computing capabilities and reconfigurability of FPGAs in particular make them a promising platform for acceleration of multiple different NN architectures~\citep{mittal2020survey}. Nevertheless, acceleration of BNNs specifically has not gained similar interests in the research community and there are only few works which approached this challenge~\citep{cai2018vibnn, myojin, 9116302}.

In \textit{VIBNN}~\citep{cai2018vibnn}, the authors developed an efficient FPGA-based accelerator for BNNs, however, they focused on BNNs consisting only of linear layers. Myojin \textit{et al.}~\citep{myojin} propose a method for reducing the sampling time required for MCD~\citep{gal2016dropout} in edge computing by parallelising the calculation circuit using an FPGA. 
However, their method needs to binarise the BNN and they again focus only on linear layers. Awano \& Hashimoto~\citep{9116302} propose a custom inference algorithm for BNNs consisting exclusively of linear layers - \textit{BYNQNet} which employs quadratic nonlinear activation functions and hence the uncertainty propagation can be achieved using only polynomial operations.
Although the design can achieve a high throughput,
the restriction of the nonlinear activation functions limits generality for different application scenarios. In~\citep{azevedo2020stochasticyolo}, the authors propose software-based intermediate-layer caching (IC), evaluated in last layer BNNs.

In comparison to these works, we focus on accelerating BNNs consisting of different layers with or without residual connections~\citep{he2016deep}, including convolutions or pooling, that have been popular in the present-day networks~\citep{fan2019f, he2016deep}. Additionally, our work wants to appeal to already wide-spread MCD without any additional software re-implementation effort.

\subsection{Monte Carlo Dropout (MCD)}\label{sec:mcd}
The concept MCD~\citep{gal2016dropout} lays in casting dropout~\citep{srivastava2014dropout} in NNs as Bayesian inference. 
Unlike the dropout used in standard NNs which is only enabled during training,
MCD applies the dropout during both training and evaluation.
MCD can be described as applying a random filter-wise mask $\boldsymbol{M}_i \in \mathbb{R}^{F_i}$ to the output feature maps $\boldsymbol{Y}_i$ of layer $i$ with $F_i$ filters.
The mask $\boldsymbol{M}_i$ follows a Bernoulli distribution $p(\boldsymbol{M}_i| p_i)$ which generates binary random variables (0 or 1) with the probability $p_i \in [0, 1]$. $p$ practically trades-off certainty, accuracy and calibration of the BNN.
After MCD removes the output feature maps with zeros,
the non-zero elements are then scaled by $\frac{1}{1-p_i}$.
To get the final output $\boldsymbol{O}_i$ under MCD,
the computation can be formulated as: $
    \boldsymbol{O}_i = \frac{1}{1-p}_i (\boldsymbol{Y}_i\odot \boldsymbol{M}_i)$
where $\odot$ represents a Hadamard product and $\boldsymbol{M}_i$ is generated by a Bernoulli sampler at runtime for different filters and layers. The uncertainty estimation and prediction is thus obtained by running the same input through the BNN $S$ times, each time with different set of sampled masks $\boldsymbol{M}$ for each layer $i$ where MCD is applied, and averaging the outputs.
The works~\citep{gal2016dropout, kendall2015bayesian} demonstrate that MCD can achieve a high quality of uncertainty estimation. 


\subsection{Partial Bayesian Inference}\label{sec:partial_bnn}
A full BNN should be trained with MCD applied after every layer~\citep{gal2016dropout}. However, the authors in~\citep{kristiadi2020being, kendall2015bayesian} have demonstrated theoretically and empirically that making a standard NN Bayesian in different parts of the NN, thus making it partially Bayesian, can improve uncertainty estimation and it can also improve accuracy.
Assuming there is an $N$-layer NN, 
partial Bayesian inference applies MCD in the last $L; L\leq N$ layers and makes the first $N-L$ layers behave as a feature extractor for the Bayesian remainder.
Partially applied dropout then represents a trade-off between hardware, algorithmic performance and uncertainty estimation~\citep{kendall2015bayesian}.
In this paper,
we exploit this trade-off by proposing a framework for exploring the positioning of MCD at different parts of the NN which results in a partial BNN.

\section{Hardware Design}\label{sec:hardware}

\begin{figure}[b]
\centering
\includegraphics[width=0.45\textwidth]{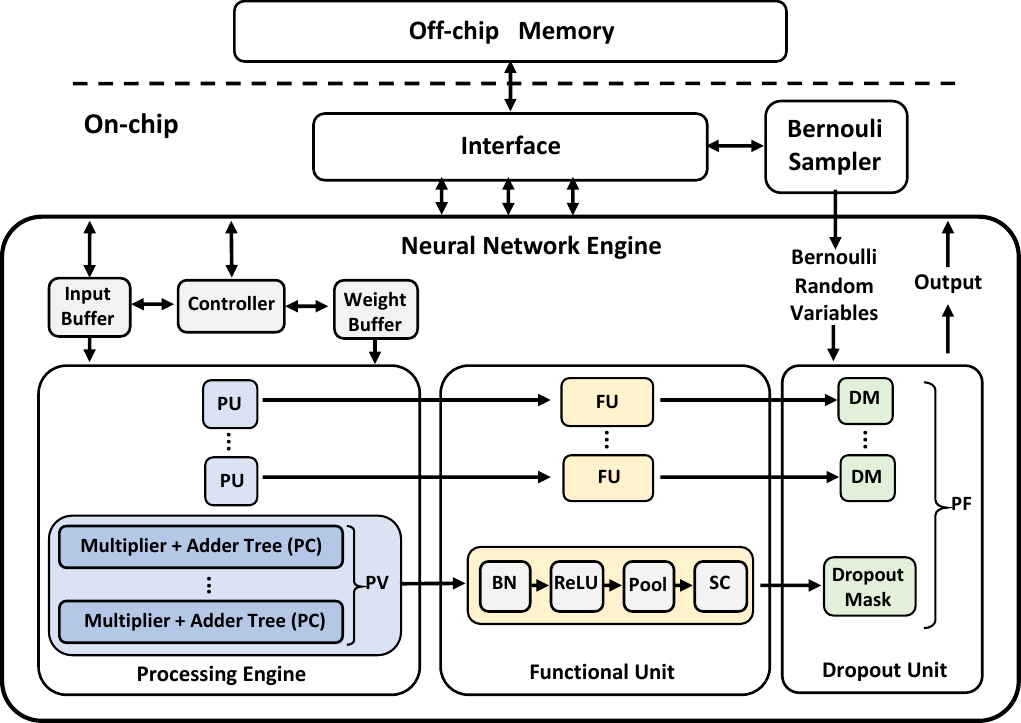}
\caption{Overview of the FPGA-based accelerator.}
\label{fig:hw_overview}
\end{figure}

\subsection{Design Overview}
An overview of the proposed hardware design is illustrated in~\figref{fig:hw_overview}.
The computation of the BNN is performed layer-by-layer using the same hardware design.
The intermediate outputs of each layer are transferred to off-chip memory to reduce the on-chip memory consumption,
and they are loaded back to the input buffer for processing of the next layer.
The weights of different layers are stored in off-chip memory and loaded to the weight buffer while processing the corresponding layer.

The main component in the proposed accelerator is a neural network engine (NNE), which is designed for running one layer at a time and general enough to run linear and convolutional layers with different kernel sizes. The NNE consists of a processing engine (PE), a functional unit (FU) and a dropout unit (DU).
These sub-modules are queried in a pipeline manner to improve the hardware performance.
The PE is designed to perform matrix multiplication,
which supports three types of fine-grained parallelism: filter parallelism ($PF$), channel parallelism ($PC$) and vector parallelism ($PV$).
In PE, there are $PF$ processing units (PUs).
Each PU contains $PV$ multiplication-addition modules and each module contains $PC$ multipliers followed by an adder tree for channel accumulation.
After each PU, there is a chain of FU modules including batch normalization (\textit{BN})~\cite{ioffe2015batch}, Rectified linear unit (\textit{ReLU}) activation, Pooling (\textit{Pool}) and Shortcut (\textit{SC}). The DU is placed at the end,
which is a batch of multiplexers controlled by the zeros and ones generated from the Bernoulli sampler.

\subsection{Bernoulli Sampler}\label{sec:hardware_dropout}
MCD is applied filter-wise, which means the number of Bernoulli random variables generated for each layer $i$ is equal to the number of output filters.
Therefore, we adopt the single-bit linear feedback shift register (LFSR) design to implement a Bernoulli sampler, which is illustrated in~\figref{fig:hw_bernoulli}.
\begin{figure}[H]
\centering
\includegraphics[width=0.35\textwidth]{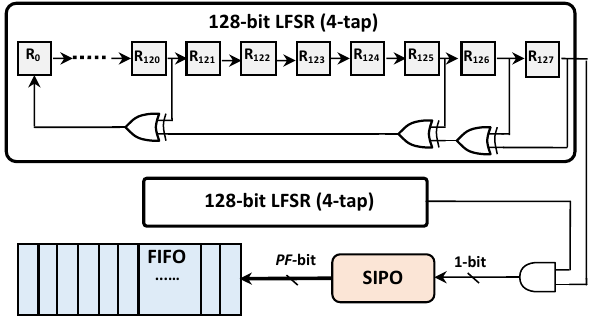}
\caption{Hardware architecture of the implemented Bernoulli sampler.}
\label{fig:hw_bernoulli}
\end{figure}
The LFSR is composed of a chain of shift registers formed as a loop.
The maximum sequence length $S_{max}$ of LFSR depends on the number of shift registers $N_{reg}$ used in the loop: $S_{max} = 2^{N_{reg}} - 1$.
The used LFSR design would take 1500 years to iterate through the whole sequence when clocked at 160MHz~\cite{andraka1998fpga}.
Since a single LFSR can only support Bernoulli sampling with $0.5$ probability,
the number of LFSRs depends on the required dropout rate.
For instance,
two LFSRs with an extra AND gate are required to implement Bernoulli sampler with $p=0.25$.
Also,
as mentioned in~\secref{sec:hardware},
the NNE only processes $PF$ filters at a time,
so we design a serial-in-parallel-out (SIPO) module, placed after LFSRs, to form a single Bernoulli bit of a $PF$-bit MCD mask.
Since different filters are processed at different speeds,
a first-in-first-out (FIFO) buffer is placed at the end of the Bernoulli sampler to cache generated Bernoulli random variables and pop out the mask when required. In case the overall processing is parallelised it is not necessary to use more than one sampler, however, the samples sampled during runtime for each instance need to be distinct.

\subsection{Intermediate-layer Caching (IC)}\label{sec:intermediate_caching}
To further improve the overall hardware performance,
we propose a hardware implementation of IC technique~\cite{azevedo2020stochasticyolo} to decrease the required compute and the number of memory accesses.
An example of using IC is illustrated in~\figref{fig:hw_ll_chache},
where the NN contains two layers and it only requires the user to apply the dropout mask and run the last layer $S$ times when the partial Bayesian technique is applied.
In IC, the input of the last layer is stored on chip until the sampling is finished.
Assuming the NN requires to run the last $L$ layers $S$ times to obtain the prediction,
the IC can reduce the compute by $(N-L) \times S$ times and the number of memory accesses by $L$ times.
\begin{figure}[t]
\centering
\includegraphics[width=0.43\textwidth]{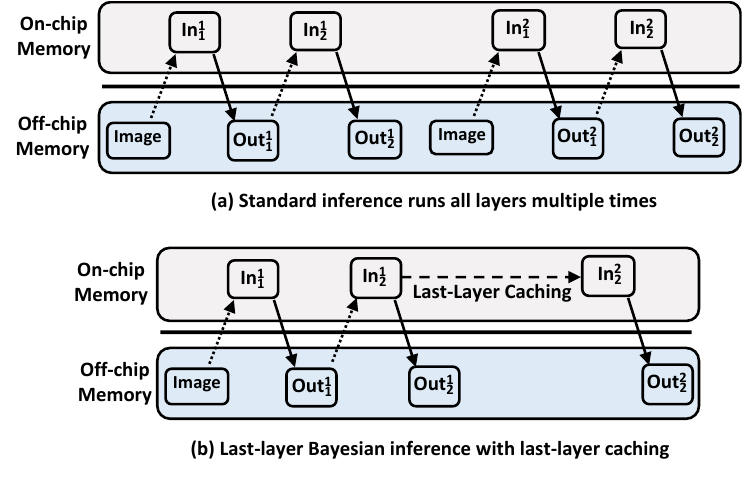}
\caption{An example of a two-layer neural network which computes two samples to obtain the prediction. The input and output data are denoted by In$^{i}_{j}$ and Out$^{i}_{j}$ where $i$ means $i$\textsuperscript{th} iteration and $j$ represents $j$\textsuperscript{th} layer.}
\label{fig:hw_ll_chache}
\end{figure}

\section{Optimization Framework}\label{sec:framework}

\subsection{Workflow of Framework}
As mentioned in~\secref{sec:partial_bnn},
partially applying MCD represents a trade-off between latency, accuracy, confidence and uncertainty estimation.
The trade-off is decided by three types of parameters: \textit{1)} $L$ which denotes the portion of Bayesian layers, \textit{2)} $S$ which represents the number of times needed to repetitively run the Bayesian parts and \textit{3)} $PC, PF, PV$ which represent hardware parallelism.
In this paper, we propose a framework, shown in~\figref{fig:framework_overview}, which automatically optimizes the configuration of the BNN with respect to parameters $L$, $S$ and $PC, PF, PV$ according to user requirements for the target hardware platform.
In our hardware design space, we consider the domains for both $PC$ and $PF$ as $\{8, 16, 32, 64, 128\}$ and $PV$ can be chosen from $\{1, 4, 8, 16\}$.

At the beginning, the framework requires users to specify the hardware constraints, optimization mode and the minimal requirement for each metric.
The hardware constraints include the available DSPs and memory resources of the target hardware platform.
The optimization mode is selected from optimal-latency, optimal-accuracy, optimal-uncertainty prediction and optimal-confidence to minimise or maximise the chosen objective through greedy optimisation with respect to software and hardware configurations. 
\begin{figure}[t]
\centering
\includegraphics[width=0.41\textwidth]{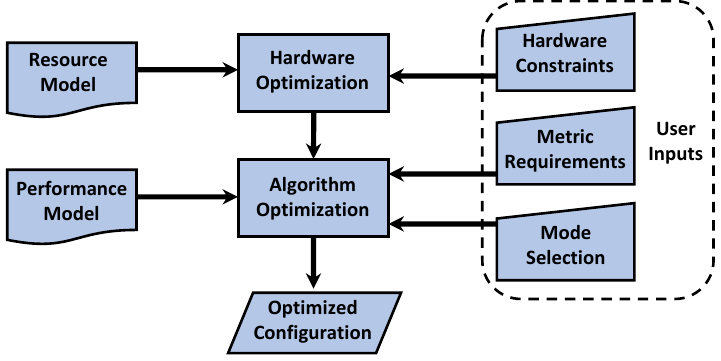}
\caption{Overview of the optimization framework.}
\label{fig:framework_overview}
\end{figure}
The first optimization is the hardware optimization, which determines the maximum parallelism level implementable on the target hardware in terms of $PC, PF, PV$.
The resource model is used at this step to estimate the resource consumption given the available degrees of parallelism.
During algorithmic optimization, based on the determined hardware parameters,
we obtain the latency from the performance lookup table for various BNNs with different $L$ and $S$.
At the same time, the accuracy, the quality of uncertainty prediction and confidence of the BNN are evaluated in software.
Then,
the configurations which do not meet the minimal requirements are filtered.
The final configurations are selected according to the optimization mode specified at the beginning.

\subsection{Resource Model}\label{sec:res_model}
As memory and DSPs are the limiting resources in FPGA-based NN accelerators~\cite{liu2018optimizing},
we mainly consider the memory and DSPs usage. The DSP usage depends on the multipliers used in the NNE.
Due to 8-bit processing,
we implement two multipliers using one DSP and thus the DSP consumption can be calculated as $DSP = \frac{PC \times PF \times PV}{2}$.
The memory resources are mainly consumed by the weight buffer, input buffer in the NNE and the FIFO buffer in the Bernoulli sampler.
As the width of the FIFO is $PF$,
its memory consumption can be represented as $MEM_{FIFO} = D \times PF \times DW$, where $D$ represents the depth of the FIFO used in the Bernoulli sampler and $DW$ is the data width.
As our design processes the NN layer-by-layer, the memory usage of input buffer is dominated by the layer with the maximal input size as $MEM_{in} = \max\limits_{i = 1, \ldots, N}(C_{i} \times H_{i} \times W_{i}) \times DW$, where $H_{i}, W_{i}$ and $C_{i}$ are the height, width and number of input channels of the $i$\textsuperscript{th} layer respectively.
Since the weight buffer only needs to cache $PF$ filters, the memory consumption of the weight buffer can be formulated as $ MEM_{weight} = \max\limits_{i = 1, \ldots, N}(C_{i} \times K_{i} \times K_{i}) \times PF \times DW$, where $K_{i}$ is the kernel size of the $i$\textsuperscript{th} layer.

\section{Experiments}\label{sec:experiments}

\begin{table*}
\centering
\caption{The resultant configurations of BNNs under different optimization modes.}
\label{tb:design_space}
\scalebox{0.96}{
\setlength\tabcolsep{6pt} 
\begin{tabular}{C{1.2cm}|C{1.9cm}|C{1.5cm}|C{1cm}| C{1cm} | C{1cm}|C{1.5cm}| C{1.5cm}| C{2cm}}
\toprule
& \multirow{3}{*}{\bf Opt-Mode} & \multirow{3}{*}{$\{\boldsymbol{L}, \boldsymbol{S}\}$} & \multicolumn{3}{c|}{{\bf Latency} [ms]  $\downarrow$} & \multirow{3}{*}{\textbf{aPE} [nats] $\uparrow$}& \multirow{3}{*}{\textbf{ECE} [\%] $\downarrow$} & \multirow{3}{*}{\textbf{Accuracy} [\%] $\uparrow$} \\ 
\cmidrule{4-6}

&  &  &FPGA  &CPU & GPU &  & & \\

\midrule
\multirow{5}{*}{\textit{LetNet-5}} 
 & {\textit{Opt-Latency}} & 1, 3 & 0.42 & 0.67 & 0.24 & $ 0.63 \pm 0.09$ & $ 0.25 \pm 0.05$ & $ 99.27 \pm 0.04$  \\
 \cmidrule{2-9}
 & {\textit{Opt-Accuracy}} & $\frac{2}{3} \times N$, 100 & 14.32 & 24.69 & 12.87 & $ 0.75 \pm 0.15$ & $ 0.13 \pm 0.03$ & $ 99.39 \pm 0.05$  \\
\cmidrule{2-9}
 & {\textit{Opt-Uncertainty}} & $N$, 100 & 14.83 & 42.0 & 19.91 & $ 1.06 \pm 0.19$ & $ 0.17 \pm 0.04$ & $ 99.32 \pm 0.04$  \\
 \cmidrule{2-9}
& {\textit{Opt-Confidence}} & $N$, 9 & 1.29 & 3.68 & 1.68 & $ 0.98 \pm 0.18$ & $ 0.1 \pm 0.04$ & $ 99.31 \pm 0.03$  \\
\midrule

\multirow{5}{*}{\textit{VGG-11}} & {\textit{Opt-Latency}} & 1, 3 & 0.57 & 0.95 & 0.68 & $ 1.38 \pm 0.28$ & $ 2.8 \pm 0.12$ & $ 95.38 \pm 0.1$  \\
\cmidrule{2-9}
 & {\textit{Opt-Accuracy}} & $N$, 100 & 57.32 & 186.24 & 88.93 & $ 1.97 \pm 0.05$ & $ 2.42 \pm 0.19$ & $ 96.49 \pm 0.05$  \\
\cmidrule{2-9}
& {\textit{Opt-Uncertainty}} & $\frac{2}{3} \times N$, 100 & 42.89 & 110.32 & 59.78 & $ 2.02 \pm 0.11$ & $ 0.41 \pm 0.05$ & $ 96.13 \pm 0.1$  \\
\cmidrule{2-9}
& {\textit{Opt-Confidence}} & $\frac{2}{3} \times N$, 100 & 42.89 & 110.32 & 59.78 & $ 2.02 \pm 0.11$ & $ 0.41 \pm 0.05$ & $ 96.13 \pm 0.1$  \\
\midrule

\multirow{5}{*}{\textit{ResNet-18}} 
& {\textit{Opt-Latency}}& 1, 3 & 0.47 & 1.31 & 0.87 & $ 0.36 \pm 0.26$ & $ 4.85 \pm 0.19$ & $ 92.84 \pm 0.16$  \\
\cmidrule{2-9}
& {\textit{Opt-Accuracy}} & 1, 8 & 0.50 & 2.03 & 1.17 & $ 0.38 \pm 0.27$ & $ 4.74 \pm 0.14 $ & $ 92.91 \pm 0.14$  \\
\cmidrule{2-9}
& {\textit{Opt-Uncertainty}} & $\frac{1}{2} \times N$, 100 & 32.04  & 173.53 & 93.23 & $ 1.27 \pm 0.27$ & $  2.74 \pm 0.31$ & $ 91.12 \pm 0.2$  \\
\cmidrule{2-9}
& {\textit{Opt-Confidence}} & $\frac{2}{3} \times N$, 3 & 1.20 & 7.66 & 3.93 & $ 1.05 \pm 0.26$ & $  1.08 \pm 0.06$ & $ 89.99 \pm 0.17$  \\
\bottomrule
\end{tabular}}
\end{table*}

\subsection{Experimental Setup}
In this paper, Intel Arria 10 SX660 FPGA is set as our target hardware platform.
1GB DDR4 SDRAM is installed as off-chip memory.
The PyTorch framework is used for the software implementation.
We focus on image classification. We evaluate the networks on tuples $(\boldsymbol{x}, \boldsymbol{y})$, where the target $\boldsymbol{y}$ is an one-hot encoding of $k=1,\ldots,K$ classes. Given the image input $\boldsymbol{x}$, we approximate the predictive distribution $p(\boldsymbol{y}| \boldsymbol{x})$ over the target with respect to $S$ samples as $\frac{1}{S}\sum_{s=1}^S p(\boldsymbol{y}| \boldsymbol{x}, \boldsymbol{M}); \boldsymbol{M}\sim p(\boldsymbol{M} | p)$, where the $\boldsymbol{M}$ is the set of Bernoulli masks and $S$ can be $S=\{3,4,5,6,7,8,9, 10, 20, 50, 100\}$. We consider $p=0.25$ for all MCD instances.

For datasets, we consider classifying images of increasing difficulty: MNIST, SVHN and CIFAR-10, through which we control the complexity of the experiments. For MNIST we implement \textit{LeNet-5}~\citep{lecun1998gradient}, \textit{VGG-11}~\citep{simonyan2014very} for SVHN and \textit{ResNet-18}~\citep{he2016deep} for CIFAR-10. We reduced the channel size of \textit{VGG-11} and \textit{ResNet-18} to fit them into memory. In terms of partial Bayesian inference, we explore adding dropout in the different parts of the NN, always following a convolutional, BN and ReLU layers, and optionally pooling. Similarly to the datasets, we explore state-of-the-art architectures of increasing complexity, whose core is widely used across practical applications. Their structural irregularities present challenges to the accelerator's design. We consider partial BNNs, such that $L=\{1, \frac{1}{3} \times N, \frac{1}{2} \times N, \frac{2}{3} \times N, N\}$. All experiments were repeated 5 times.

In addition to measuring the classification accuracy, we establish metrics for the evaluation of the predictive uncertainty and confidence. For the input that should rightfully confuse the net, we measure the quality of the uncertainty prediction with respect to random Gaussian noise with mean and variance of the training data with the average predictive entropy (aPE) over a dataset of size $E$ as: $ \textrm{aPE} =  \frac{1}{E} \sum_{e=1}^E  - \sum_{k=1}^Kp(y_{e}^k| \boldsymbol{x}_{e})\log p(y_{e}^k| \boldsymbol{x}_{e})$. Additionally, we measure the confidence with which the net is making its predictions on the test data through the expected calibration error (ECE)~\citep{guo2017calibration}. ECE signals that a BNN is uncalibrated if it is making predictions whose confidence are not matching its accuracy. We calculate ECE with respect to 10 bins.

We implement our design using Verilog and
Quartus 17 Prime Pro is used for synthesis and implementation.
Based on the resource model and the available resources on our FPGA,
$PC, PF$ and $PV$ are set to be $64$, $64$ and $1$ respectively and the final design is clocked at $225$ MHz.
The resource usage of the proposed accelerator is presented in~\tabref{table:res}. Since our accelerator is based on 8-bit precision,
the 8-bit linear quantization~\cite{jacob2018quantization} is applied on the trained models.
\begin{table}[t]
\centering
\caption{Resource utilization of the accelerator on the FPGA.}
\scalebox{0.92}{
\setlength\tabcolsep{6pt}  
\begin{tabular}{c | c | c | c| c} 
\toprule
\textbf{Resources} & \textbf{ALMs} & \textbf{Registers}& \textbf{DSPs} &\textbf{M20K} \\ 
\midrule
Used & 303,913 & 889,869 & 1,473 & 2,334 \\
\midrule
Total & 427,200 & 1,708,800 & 1,518 & 2,713\\
\midrule
Utilization & 71\% & 52\% & 97\% & 86\% \\
\bottomrule
\end{tabular}
}
\label{table:res}
\vspace{-1.5mm}
\end{table}

\subsection{Hardware Performance Comparison}
For each network, we measure the hardware performance on the FPGA, Intel Core i9-9900K CPU and NVIDIA RTX 2080 SUPER GPU, the batch size is 1 for all the hardware platforms for a fair comparison\footnote{Since PyTorch does not support 8-bit quantization on a GPU, the latency of GPU is estimated by dividing its floating-point performance by 4 times, which is the theoretically the lowest latency that the GPU can achieve.}.
For the FPGA implementation, we measure the latency with and without IC (\secref{sec:intermediate_caching})
to demonstrate its effect and the results are shown in~\tabref{tb:hardware_comparison}, the down and up arrows indicate the desired tendency for a given metric.
While comparing FPGA implementations with and without IC on \textit{VGG-11} and \textit{ResNet-18}, it can be seen that the speed up brought by IC goes down when $L$ increases and the $S$ decreases.
In comparison to CPU and GPU implementations, the BNNs on the FPGA with IC can achieve up to $15$ times and $8$ times speed up respectively.
There are two reasons for the speedup: \textit{1)} The adoption of IC technique together with MCD and partial Bayesian inference, which decreases the amount of memory accesses and computation; \textit{2)} The support for fine-grained parallelism on the accelerator, which fully utilized the extensive concurrency exhibited in BNNs.
On \textit{LeNet-5}, since the execution time is mainly occupied by the last layer, IC does not bring too much improvement on FPGA compared with GPU and CPU.
However, because the current state-of-the-art NNs only spend a small portion of the execution time in the last layer,
our accelerator can still achieve speed up in most of NNs, and thus is practical enough for real-life applications.

\begin{table}[t]
\centering
\caption{Hardware comparison between FPGA, CPU and GPU.}
\label{tb:hardware_comparison}
\scalebox{0.88}{
\setlength\tabcolsep{6pt} 
\begin{tabular}{C{1.2cm}|C{1.3cm}|C{0.65cm}| C{0.8cm}| C{0.78cm} | C{0.78cm}}
\toprule
&  \multirow{4}{*}{$\{\boldsymbol{L}, \boldsymbol{S}\}$} & \multicolumn{4}{c}{{\bf Latency} [ms] $\downarrow$} \\ 
\cmidrule{3-6}
& & \multicolumn{2}{c|}{FPGA}& \multirow{2}{*}{CPU} & \multirow{2}{*}{GPU} \\
\cmidrule{3-4}
& & w/ IC & w/o IC & &  \\
\midrule

\multirow{2}{*}{\textit{LetNet-5}} 
& 1, 100 & 13.73 & 14.38 & 11.17 & 5.81  \\
 \cmidrule{2-6}
& $\frac{2}{3}\times N$, 50 & 7.16 & 7.20 & 12.02 & 6.07   \\
\midrule

\multirow{2}{*}{\textit{VGG-11}} & 1, 100 & 0.76 & 57.3 & 11.76 & 6.33   \\
\cmidrule{2-6}
 & $\frac{2}{3}\times N$, 50 & 21.52 & 28.67 & 55.94 & 30.09   \\
\midrule

\multirow{2}{*}{\textit{ResNet-18}} 
& 1, 100 & 1.22 & 44.97 & 13.96 & 7.05   \\
\cmidrule{2-6}
& $\frac{2}{3}\times N$, 50 & 18.90 & 22.48 & 131.41 & 65.9  \\
\bottomrule
\end{tabular}}
\end{table}

We also compare our work with the other BNN accelerators~\citep{cai2018vibnn, 9116302} in~\tabref{table:sota_bnn}.
Because the three-layer BNNs evaluated in~\citep{cai2018vibnn} and~\citep{9116302} are unrealistic for real-life applications,
we run a commonly-used ResNet-101~\citep{he2016deep} on our design with MCD applied onto every layer, such that $L=N$.
However, as both~\citep{cai2018vibnn} and~\citep{9116302} do not support ResNet-101, their performance reported is still based on the three-layer BNN in their original papers.
For a fair comparison, we evaluate all the accelerators in terms of throughput, compute and energy efficiency\footnote{The energy efficiency is quoted in in giga-operations per second per watt (GOP/s/W) and the total board power consumption is 45W.}.
As shown in~\tabref{table:sota_bnn},
our accelerator can achieve 3 times to 4 times higher energy efficiency and 6 times 9 times better compute efficiency.
Also, it is worth to mention that previous BNN accelerators only support linear layers,
while our proposed accelerator is versatile enough to support a wide range operations including convolution, pooling or residual addition.

\begin{table}[t]
\centering
\caption{Performance comparison with other BNN accelerators.}
\scalebox{0.86}{
\setlength\tabcolsep{6pt}  
\begin{tabular}{c | c | c | c} 
\toprule
& \textit{VIBNN}~\citep{cai2018vibnn} & \textit{BYNQNet}~\citep{9116302}& \textit{Our work}  \\ 
\midrule
\multirow{2}{*}{\textbf{FPGA}} & Cyclone V & Zynq & Arria 10 \\
 & 5CGTFD9E5F35C7 & XC7Z020 & SX660\\
\midrule
\textbf{Clock frequency} [MHz] & 212.95 & 200 & 225\\ 
\midrule
\textbf{Total number of DSPs} & 342 & 220 & 1473\\
\midrule
\textbf{Energy} [W] $\downarrow$ & 6.11 & 2.76 & 45.00\\
\midrule
{\textbf{Throughput} [GOP/s] $\uparrow$} & 59.6 & 24.22 & 1590  \\
\midrule
{\textbf{Energy Eff.} [GOP/s/W] $\uparrow$}  & 9.75 & 8.77 & 33.3 \\
\midrule
{\textbf{Comp. Eff.} [GOP/s/DSP] $\uparrow$} & 0.174 & 0.121 & 1.079\\
\bottomrule
\end{tabular}
}
\label{table:sota_bnn}
\end{table}


\subsection{Effectiveness of Framework}
As introduced in~\secref{sec:framework},
our framework is designed to explore the trade-off between accuracy, latency, uncertainty estimation and confidence.
This section investigates design space exploration with and without user constraints.

\subsubsection{Exploration Without Constraints}
To find the global optimal latency, accuracy, uncertainty and confidence points, we set four different optimization modes: \textit{Opt-Latency}, \textit{Opt-Accuracy}, \textit{Opt-Uncertainty} and \textit{Opt-Confidence}, for all BNNs without any constrains.
The results are illustrated in~\tabref{tb:design_space}. 
The lowest latencies that our accelerator can achieve on these three NNs are $0.42$ms, $0.57$ms and $0.47$ms respectively.
With the \textit{Opt-Accuracy} mode enabled,
these three NNs can achieve $99.39$\%, $96.49$\% and $92.91$\% accuracy respectively on their corresponding datasets.
The framework also suggests different $\{L, S\}$ configurations to achieve the optimal aPE and ECE.

\subsubsection{Exploration With Constraints}
To demonstrate that our framework is able to find the optimal points when the user's requirements are given,
we set latency, accuracy and uncertainty constraints for \textit{ResNet-18} on CIFAR-10 dataset and select the \textit{Opt-Confidence} mode for optimization.
\figref{fig:dse} shows all the candidate points with respect to accuracy, latency, aPE and ECE.
The global optimal points with respect to different metrics are highlighted by the black arrows.
The feasible design space constructed by accuracy, latency and uncertainty constraints is represented by the black box.
Within this feasible design space,
our framework generates the point with the lowest ECE, which is marked by the red arrow.
Therefore, the proposed framework is able to find the optimal points when users' constraints are given.

\begin{figure}[t]
\centering
\includegraphics[width=0.83\linewidth]{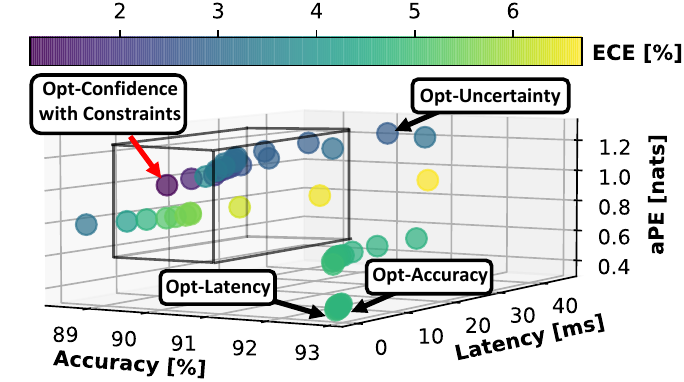}
\caption{Design space exploration with latency, accuracy and uncertainty constraints for \textit{ResNet-18} on CIFAR-10.}
\label{fig:dse}
\end{figure}


\section{Conclusion}\label{sec:conclusion}

This work proposes a high-performance FPGA-based design to accelerate Bayesian neural networks (BNNs) inferred through Monte Carlo Dropout. The accelerator is versatile enough to support a variety of Bayesian neural networks and it achieves up to 4 times higher energy efficiency and 9 times better compute efficiency than other state-of-the-art accelerators. Additionally, we presented a framework to automatically trade-off both hardware and algorithmic performance, given hardware constraints and algorithmic requirements.
In future we aim to explore neural architecture search on BNN, and co-develop the hardware design for BNNs found.
\section*{Acknowledgment}
Martin Ferianc’s conference registration and travel were sponsored by G-Research’s Small Monthly Grants Scheme and Tatra Bank's Student Grant Scheme and his work was sponsored by a scholarship from the Institute of Communications and Connected Systems at UCL.




%
\printbibliography

\end{document}